\documentclass[sigconf,dvipsnames,authorversion]{acmart}
\pdfoutput=1

\usepackage[utf8]{inputenc}
\usepackage[T1]{fontenc}

\usepackage{amssymb}        
\usepackage{balance}
\usepackage{booktabs}       % For formal tables
\usepackage{xcolor}
\usepackage{enumitem}       % fixes overfull hboxes in description envs
\usepackage[toc,acronym,nowarn]{glossaries}
\usepackage{graphicx}
\usepackage[all]{hypcap}
\usepackage{hyperref}
\usepackage{lipsum}
\usepackage{listings}
\usepackage{multirow}
\usepackage{microtype}
\usepackage{nicefrac}
\usepackage{pbox}
\usepackage{pgfplots, pgfplotstable}
\usepackage{textcomp}
\usepackage{subcaption}
\usepackage{tikz}
\usepackage{xspace}

\newacronym{OS}{OS}{Operating System}

\newacronym{IoT}{IoT}{Internet of Things}
\newacronym{ISA}{ISA}{Instruction Set Architecture}
\newacronym{SPI}{SPI}{Serial Peripheral Interface}

\makeglossaries
\glsdisablehyper % disable hyperlinks to nowhere...
\newcommand*{\eg}{e.g.,\@\xspace}

\newcommand{\esp}{ESP8266\xspace}
\newcommand{\esps}{ESP8266s\xspace}
\newcommand{\mechanism}{Harzer Roller\xspace}

\graphicspath{{graphics/}}
\DeclareGraphicsExtensions{.pdf,.jpeg,.png}

\newcommand{\Title}{Harzer Roller: Linker-Based Instrumentation for Enhanced Embedded Security Testing}

\def\etal{et al.\@\xspace}

\lstdefinelanguage{pasm}
{
	keywords=[1]{
		call,
		store,
		move,
		jump,
		inc,
		clear,
		return,
		cmp,
		jumpz,
		dec,
		hlt
	},
	keywordstyle=[1]{\color{blue}\bfseries},
	keywords=[2]{a0, a1, a2, a3, a4, a5, a6, lr},
	keywordstyle=[2]{\color{purple}},
	keywords=[3]{returnstack},
	keywordstyle=[3]{\color{RedOrange}},
	sensitive=false, % keywords are not case-sensitive
	morecomment=[l]{//}, % l is for line comment
	morecomment=[s]{/*}{*/}, % s is for start and end delimiter
	morestring=[b]" % defines that strings are enclosed in double quotes
}
\lstnewenvironment{pseudoasm}
{\lstset{
		language=pasm,
		breaklines=true,
		extendedchars=true,
		frame=none,
		showstringspaces=false,
		showtabs=false,
		numbers=none
}}
{}

\lstdefinelanguage{xtensaasm}
{
	alsoletter={.},
	keywords=[1]{
		rsr.epc1,
		l32r,
		sub,
		beqz,
		addi.n,
		l32i.n,
		l32r,
		callx0,
		s32i.n,
		wsr.epc1,
		rfe
	},
	keywordstyle=[1]{\color{blue}\bfseries},
	keywords=[2]{a0, a1, a2, a3, a4, a5, a6, lr, a15},
	keywordstyle=[2]{\color{purple}},
	keywords=[3]{canary, stack_chk_f, returnstack, stack_ptr, good},
	keywordstyle=[3]{\color{RedOrange}},
	sensitive=false, % keywords are not case-sensitive
	morecomment=[l]{//}, % l is for line comment
	morecomment=[s]{/*}{*/}, % s is for start and end delimiter
	morestring=[b]" % defines that strings are enclosed in double quotes
}
\lstnewenvironment{xtensa}
{\lstset{
		language=xtensaasm,
		breaklines=true,
		extendedchars=true,
		frame=single,
		showstringspaces=false,
		showtabs=false,
		numbers=left,
		linewidth=7.85cm
}}
{}
\lstnewenvironment{xtensar}
{\lstset{
		language=xtensaasm,
		breaklines=true,
		extendedchars=true,
		frame=single,
		showstringspaces=false,
		showtabs=false,
		numbers=right,
		linewidth=7.85cm
}}
{}

\begin{document}
\newsavebox\pasmcall
\begin{lrbox}{\pasmcall}
	\begin{pseudoasm}
...
...
...
call fct
...
...
	\end{pseudoasm}
\end{lrbox}
\newsavebox\hrpseudo
\begin{lrbox}{\hrpseudo}
	\begin{pseudoasm}
fct:
inc returnstack
move =returnstack, lr
move lr, 0xabababab
jump hr_fct
	\end{pseudoasm}
\end{lrbox}
\newsavebox\protfunpseudo
\begin{lrbox}{\protfunpseudo}
	\begin{pseudoasm}
hr_fct:
...
...
...
...
return
	\end{pseudoasm}
\end{lrbox}
\newsavebox\illhandlerpseudo
\begin{lrbox}{\illhandlerpseudo}
	\begin{pseudoasm}
illegal_instruction:
  cmp a0, 0xabababab
  jumpz fail
  clear // clear exc.
  move a1, =returnstack
  dec returnstack
  jump a1

  fail:
  call print_fail
  hlt
	\end{pseudoasm}
\end{lrbox}

\copyrightyear{2019}
\acmYear{2019}
\setcopyright{acmlicensed}
\acmConference[ROOTS'19]{Reversing and Offensive-oriented Trends Symposium}{November 28--29, 2019}{Vienna, Austria}
\acmBooktitle{Reversing and Offensive-oriented Trends Symposium (ROOTS'19), November 28--29, 2019, Vienna, Austria}
\acmPrice{15.00}
\acmDOI{10.1145/3375894.3375897}
\acmISBN{978-1-4503-7775-1/19/11}

\title[Harzer Roller]{\Title}

 \author{Katharina Bogad}
 \affiliation{
   \institution{Fraunhofer AISEC}
   \city{Garching near Munich}
   \state{Germany}
 }
 \email{katharina.bogad@aisec.fraunhofer.de}
 
 \author{Manuel Huber}
 \orcid{0000-0003-0829-6902}
 \affiliation{
   \institution{Fraunhofer AISEC}
   \city{Garching near Munich}
   \state{Germany}
 }
 \email{manuel.huber@aisec.fraunhofer.de}

% acm
\begin{abstract}
Due to the rise of the Internet of Things, there are many new chips and platforms available for hobbyists and industry alike to build smart devices. The SDKs for these new platforms usually include closed-source binaries containing wireless protocol implementations, cryptographic implementations, or other library functions, which are shared among all user code across the platform. Leveraging such a library vulnerability has a high impact on a given platform. However, as these platforms are often shipped ready-to-use, classic debug infrastructure like JTAG is often times not available.

In this paper, we present a method, called \mechanism, to enhance embedded firmware security testing on resource-constrained devices.
With the \mechanism, we hook instrumentation code into function call and return.
The hooking not only applies to the user application code but to the SDK used to build firmware as well.
While we keep the design of the \mechanism generally architecture independent, we provide an implementation for the \esp Wi-Fi IoT chip based on the xtensa architecture.

We show that the \mechanism can be leveraged to trace execution flow through libraries without available source code and to detect stack-based buffer-overflows.
Additionally, we showcase how the overflow detection can be used to dump debugging information for later analysis.
This enables better usage of a variety of software security testing methods like fuzzing of wireless protocol implementations or proof-of-concept attack development.

\end{abstract}



\maketitle

\section{Introduction}

In recent years, significant advances have been made in embedded technology. 
Cheap computing technology paired with pervasive internet connectivity
led to the rise of a new class of embedded devices and computing technologies, which have been summarized
under the name \gls{IoT}.

Historically, these highly embedded computing devices have a bad track record
regarding their security and have been a prime target for hackers of all sorts \cite{mirai-iot-attacks}. Oftentimes
these devices are built to fulfil a designated task and either employ no or insufficiently secure means to upgrade the embedded
firmware. Although this is starting to change, the firmware life cycle poses a major challenge.
As these devices are often spatially dispersed and deployed in large quantity,
the firmware needs to be especially secure. Any attack that could reliably achieve execution of arbitrary code
also has the ability to permanently destroy deployed hardware, for example through targeted \gls{SPI} flash wear-out of the
boot sector. A common off the shelf \gls{SPI} flash memory chip, for instance, the Winbond W25Q128V, is specified with ,,more than 100,000`` program and erase cycles \cite{winbond-flash}.
When targeted at a relatively small number of flash cells, flash wearing can be achieved in a few minutes.
Depending on the specific device in question, other DoS attacks or device misuse might be possible as well \cite{esp-hacks}.

In the past, \gls{IoT} devices have been leveraged for large scale attacks, for example with the Mirai botnet \cite{mirai-iot-attacks}. Efforts have been made to secure
the development process of firmware, however little attention has been drawn on coprocessors of
such an \gls{IoT} system. A broadly used communication/Wi-Fi coprocessor with over 100 million
devices in use across the globe \cite{esp-sales} is the \emph{\esp} \cite{esp-datasheet} family of the Chinese company
Espressif, or its successor, the \emph{ESP32} \cite{esp32-datasheet}. These chips either employ a standard AT firmware which provides network connectivity
over a serial UART connection or can alternatively be programmed with custom firmware using
Espressif's SDK.
While the SDK contains some open source components, various BLOBs linked into every
built firmware remain. There have been efforts to reverse engineer these parts \cite{esp-open-sdk} to establish a fully open source stack, but some of these BLOBs are still required.

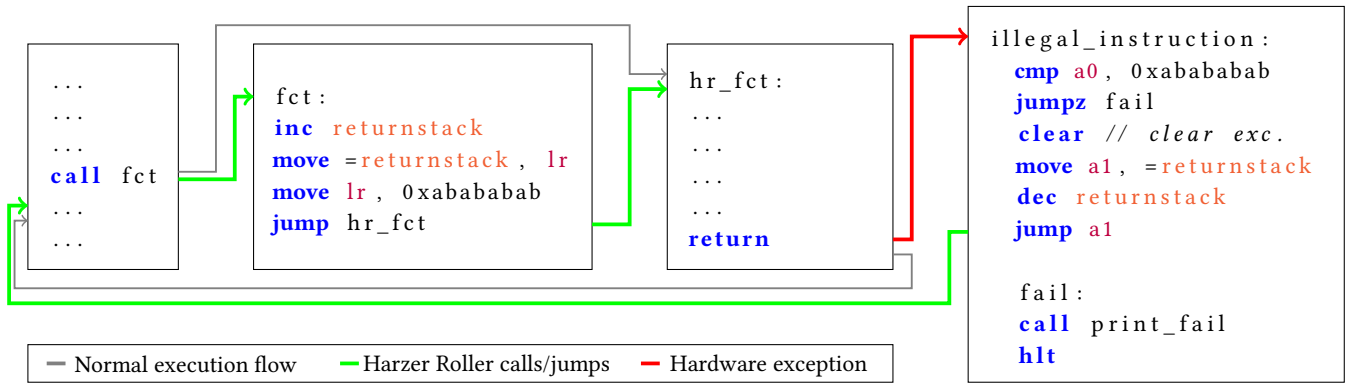
\begin{figure*}[t]
	\begin{tikzpicture}
	%%%%%%%%%%%%%%%%%%%%%%%%%%%
	%% BASIC BLOCKS
	%%%%%%%%%%%%%%%%%%%%%%%%%%%
	\draw (0, 1.5) rectangle (2, 4.5) node [pos=.5] {\parbox{1.5cm}{
		\usebox\pasmcall
	}};
	\draw (3, 1.5) rectangle (7.5, 4.5) node[pos=.5] {\parbox{4cm}{
		\usebox\hrpseudo
	}};
	\draw (11.5, 1.5) rectangle (8.5, 4.5)  node[pos=.5] {\parbox{2.5cm}{
			\usebox\protfunpseudo
	}};
	\draw (17.5, 0) rectangle (12.5, 5)  node[pos=.5] {\parbox{4.5cm}{
			\usebox\illhandlerpseudo
	}};
	%%%%%%%%%%%%%%%%%%%%%%%%%%%%%
	%% Control flow
	%%%%%%%%%%%%%%%%%%%%%%%%%%%%%
	% normal, without harzer roller
 	\draw [line width=0.25mm, gray, ->] (2, 2.8) -- (2.5, 2.8) -- (2.5, 4.75) -- (8, 4.75) -- (8,4.1) -> (8.5, 4.1);
	\draw [line width=0.25mm, gray, ->] (11.5, 1.7) -- (11.75, 1.7) -- (11.75, 1.5) -- (11.75, 1.25) -- (0, 1.25) -- (-0.1775, 1.25) -- (-0.1775, 2.15) -> (0, 2.15);
% 	harzer roller normal calls
 	\draw [line width=0.5mm, green, ->] (2, 2.7) -- (2.75, 2.7) -- (2.75, 3.8) -> (3, 3.8);
 	\draw [line width=0.5mm, green, ->] (7.5, 2.1) -- (8, 2.1) -- (8, 3.9) -> (8.5, 3.9);
	\draw [line width=0.5mm, green, ->] (12.5, 2) -- (12.25, 2) -- (12.25, 1.05) -- (0, 1.05) -- (-0.25, 1.05) -- (-0.25, 2.35) -> (0, 2.35);
% 	IllegalInstruction path
	\draw [line width=0.5mm, red, ->] (11.5, 1.9) -- (11.75, 1.9) -- (11.75, 4.6) -> (12.5, 4.6);
	%%%%%%%%%%%%%%%%%%%%%%%%%%%%%
	%% Legende
	%%%%%%%%%%%%%%%%%%%%%%%%%%%%%
	\draw(0, 0) rectangle(11.5, 0.5);
	\draw [line width=0.5mm, gray] (0.25, 0.25) -- (0.5, 0.25);
	\draw (2.1, 0.25) node {Normal execution flow};
	\draw [line width=0.5mm, green] (4.15, 0.25) -- (4.4, 0.25);
	\draw (6.1, 0.22) node {\mechanism calls/jumps};
	\draw [line width=0.5mm, red] (8.15, 0.25) -- (8.4, 0.25);
	\draw (9.85, 0.22) node {Hardware exception};
	\end{tikzpicture}
	\caption{Call and return control flow hijacking with the \mechanism in pseudo-assembly.}
	\Description[Control flow with and without \mechanism binary instrumentation.]{Control flow with and without \mechanism binary instrumentation in pseudo-assembly. Harzer roller hijacks the execution flow through linker-based instrumentation enabling to detect memory overwrites on the stack.}
	\label{cf_hijacking_fig}
\end{figure*}

The result is that even if the main application code is correct and free of bugs, the SDK could potentially
introduce vulnerabilities that are highly invisible due to the nature of binary code. While some vendors open source their SDKs -- or at least make it available under a source available license -- often times, the SDK is provided in binary-only form; thus preventing the use of source code instrumentation techniques. To actively search for vulnerabilities in such a scenario, black box testing of these closed-source components needs to be conducted. However due to the constraints
of embedded systems, usually there is no MMU available. This has consequences unfavourable for vulnerability testing:
(1) there is no means to intercept memory accesses and (2) most of the times there is no way to trap accesses to unmapped memory (such an access either returns a repeated pattern of fixed values, completely random values, zero or any combination of those).
This makes memory corruptions highly invisible, as pointed out by Muench \etal \cite{EURECOM+5417}.
For instance, memory corruptions might not immediately trigger a fault but only become obvious at a later crash or malfunction of the device.
Furthermore, source code instrumentation using stack
canaries or various sanitizers is usually neglected to save the limited computational resources.

To increase visibility of memory corruptions, happening for example through stack overflows, various approaches have been proposed before:
(1) full system virtualization, where an \gls{IoT} chip in its entirety is replicated in a virtual machine like in \cite{esp-qemu}, which is oftentimes difficult to achieve because the exact environment
and all physical interactions (with SPI flash, GPIO pins, ...) are hard to simulate, (2) partial virtualization like PROSPECT \cite{Kammerstetter:2014:PPP:2590296.2590301}
which solves this problem by forwarding hardware and pin access to the real device but impacts execution speed or traditional debug interfaces like the industry-standard JTAG port, which is -- especially in commercially available chips -- oftentimes not available. Depending on the exact virtualization setup, having access to the source code might be a requirement.

In this paper, we present a novel method to instrument object-only code that is commonly found in SDKs at the linker stage. We present an approach to instrument function calls as well as returns from these functions, using only means provided by the linker. Our approach is feasible even without source code access of the object files to be linked.
We showcase our method on an example platform, the Espressif \esp based on the xtensa \gls{ISA}, and show function call tracing and detection of overwritten saved return addresses as two possible real-world applications of the our method, as well as their application in automated vulnerability discovery via e.g. fuzzing.
Inspired by stack canaries, we call our method \emph{\mechanism} \cite{doi:10.1002/ciuz.201500708}.
Furthermore, we discuss potential attack vectors specific to the \esp.

\section{Linker-based Instrumentation} \label{lbi}

In this section, we describe the design of the \mechanism. We first give a brief overview, continue by explaining the \mechanism's call-path instrumentation, and, based on this technique, introduce return-path instrumentation.

We design the \mechanism in a way that allows us to hook the control flow of executables on calls and returns to and from subroutines, respectively. To achieve this,
we modify the object files before linking the final executable so that the linker relocates calls to an instrumented function to an assembly sequence of our choice (see \autoref{sec:instr:cpi}),
instead of modifying the pro- and epilogues of the function themselves. At the end of our injected assembly sequence, we jump to the called function. Our injected assembly sequence may override the link register with a canary value that is guaranteed to result in an illegal instruction exception when jumped to. We introduce a fault handler that can recover from such an exception, thus hooking into returns from subroutines (see \autoref{sec:instr:rpi}). We ensure that our method is able to preserve semantics between function calls and returns.

\autoref{cf_hijacking_fig} depicts this linker-based binary instrumentation which we explain in the following in more detail.
It is noteworthy that the instrumentation of the return path is optional and the \mechanism can also be used to instrument the call path only.

\subsection{Call-Path Instrumentation}
\label{sec:instr:cpi}

In our approach, we treat functions as black boxes. We thus need to preserve all registers and stack contents throughout the instrumentation. These are considered input to a function. A special case to this is the link register which should never be modified by the called subroutine.\footnote{Whenever we refer to the link register,
we mean the place the saved return address is stored. While on some architectures a CPU register is used, the return address may be placed on the stack on other architectures.}
Modifying the contents of this link register however is the target of an attacker attempting to gain control over the execution flow.

To instrument any function that gets called by either another object or the user application, we need to wrap the function and inject code into the execution flow.
For function entry (i.e. upon a call to a function), we leverage \texttt{gcc}'s \texttt{-ffunction-sections} flag, which places every function in a source file in its own section. While our method also works without this flag, it is
more effective when enabled because we have a higher number of relocations to hijack. We inject the code into the execution flow in three steps: First, we rename the symbol
in the object file that should be instrumented and prefix it with some value, for example \texttt{hr\_}. Obviously, we need to make sure that this renaming does not introduce
conflicts with already existing symbols in the object files to be linked. We then re-add all symbols we renamed in the first step, but instead of pointing to the concrete implementation provided
in the object file, we leave them as \texttt{UNDEFINED} in order to be filled in by the linker later. If the function gets called from another object file, this shifts the execution flow from the grey direct call in the left half of \autoref{cf_hijacking_fig}. Last, we look for all references to the re-named functions in all sections with relocation
information and rewrite those references that point to the renamed symbols to point to the added imports instead. In this way, we shift the calls within the object file to the \mechanism control flow path. We then generate another object file that contains all wrapping code needed for the instrumented object file and finally ensures the jumps to the real, renamed function.
This completes the upper green call path depicted in \autoref{cf_hijacking_fig}. To generate a valid executable file, we link both object files: the target we modified and the file containing our injected code.

\subsection{Return-Path Instrumentation}
\label{sec:instr:rpi}

So far, we injected code into the call path (left half of \autoref{cf_hijacking_fig}). To be able to inject code into the return path too, we ensure that the \mechanism catches all return paths.
For example, it is not uncommon to have a function check its input at the beginning and return early when it detects invalid input. These return statements do not necessarily need to be in the same basic block. Therefore, any function can have any finite number of return points.
In our method, we rely on the fact that some instructions on a microprocessor are not valid and guaranteed to cause an exception, which may
be handled by an exception handler. We can use this to craft a special return address that will generate such an illegal instruction exception and continue to do so
even if that return address is partially overwritten (e.g. by placing the target address into unmapped memory, which usually generates an exception when jumped to).

We use the call-path instrumentation described in the previous section to modify the link register within the injected wrapper function before control is passed to the instrumented function. As depicted in \autoref{cf_hijacking_fig}, we overwrite the contents of the link register with a special canary value, that is chosen in a way to guarantee an invalid instruction exception to occur when jumped (or returned) to. As depicted with a red arrow in \autoref{cf_hijacking_fig}, this shifts execution flow from the direct return into the exception instrumentation path.

Since we are able to inject code before the execution flow shifts to the called function and also on every return path, we can now transparently modify input to the function
and validate the functions output and/or rewrite it. To pass control flow back to the function that called our \mechanism instrumented function, we need to restore the state of execution. This mandates two requirements: first, all instrumentation in the return path must not overwrite any registers containing return values, callee-saved registers or stack contents still in use. Second, because the link register gets overwritten to be able to hook this return path, we need to restore the address that the return should jump to. We do this by saving the link register to a memory region not used by either stack or heap before overwriting it. We call this structure \emph{return stack} as it is a FIFO queue of saved return addresses. Depending on the concrete instrumentation done in this step, we could also save additional metadata to this structure that is only known upon call, but needed in the return path. An example of such data would be the name of the called function.

In the following section, we provide a proof of concept implementation of the \mechanism. While it is targeted at a specific platform, our method is generally applicable to other platforms as well.

\section{Implementation on the ESP8266} \label{sec:impl}
For a showcase implementation of our concept we chose Espressif's \esp chip as a platform. Its MIT-licensed \cite{sdk-license} SDK, which is available online \cite{sdk-github}, contains numerous BLOBs to dissect. Specifically, we ran our experiments with firmware built on top of version 3.0 (\texttt{2f9e0bb}) of the \esp NONOS SDK.

The \esp is based on the Tensilica xtensa ISA family \cite{esp-datasheet}. As the ISA is highly customizable, it can be configured with or without certain features like MMU, JTAG or various DSPs. The exact configuration of the architecture for the \esp processor is unknown, however the general consensus seems to be that very few features above the base package are included; especially no JTAG or MMU features \cite{crosstool-esp-config}. 
We first describe our implementation for call path instrumentation and function wrapping.
We then elaborate on how to obtain the exception table on the \esp.
We require this table for the instrumentation of the return path, which we describe subsequently.

\subsection{Call Path Instrumentation and Function Wrapping}
Because only limited space is available on the SPI flash chips supported by the \esp SDK, we must take care to keep the overhead of function wrapping as low as possible. In the xtensa ISA, we do this by taking advantage of the narrow-encoded instructions, like \texttt{addi.n}, which only take up two instead of three bytes.

We also observe that only a small portion of the actual wrapping code depends on the function that is  called. We can thus save additional space by separating the wrapper code into an independent part that we only put once in the resulting firmware file and by generating as less instructions as possible for each wrapped function. Of course, the resulting size depends on the functionality that should be achieved with the instrumentation, however we assume that most of this size cost can be located in the function independent part of the handler.

As the \mechanism injects itself into function calls, it must be completely transparent to the caller and the callee (with the obvious exception being the return address); in particular once the return address is saved, the stack pointer and all registers except \texttt{a0} -- the link register of xtensa -- must be the same as without instrumentation. This requires space in memory to perform these operations. We solve this problem by allocating a temporary stack frame to save all used callee-save registers.

Still, we need one caller-save register to hold the address of the wrapped function. We chose \texttt{a15} as this is inherently a caller-save register. Therefore, it can be modified freely by the called function, which is guaranteed not to depend on this register.

The second part of the wrapping code is relatively straight forward: we save the actual return address and the associated information to the current cell of the return stack, increase the pointer to the return stack, restore the registers \texttt{a1} to \texttt{a3} and jump to the called function. We set \texttt{a0} to the canary value, in our case 
\texttt{0xdeaddead}. While the exact value of the canary is not important, we need to choose it in a way such that we can guarantee that it reliably generates an exception. We settled on this particular value because it has the added benefit of being very easy to spot in a debugger or memory dump and rather unlikely to be incidentally encountered. 

\subsection{Registering the Exception Handler}
To make use of the return-path instrumentation, we need to register our own exception handler for illegal instruction exceptions. Unfortunately, there is no documented way of doing this with an API function of the SDK and neither the SDK API Reference \cite{esp8266-sdk-reference} nor the architecture manual \cite{xtensa-isa} specify the mechanism by which exceptions are actually handled in the \esp core.
However, the ISA manual specifies that any implementation of the exception option needs to specify a user, kernel and double exception vector. Fortunately, these vectors are specified in the
default linker script for the platform \cite{esp-exception-ld} as \texttt{0x40000030} for the kernel and \texttt{0x40000050} for the user vector, respectively.

From the memory map \cite{esp-memory-map}, we deduct that this location is in the processor's internal ROM which cannot be written to. Any code that resides in this ROM is the same across
all \esp devices with the same revision. Unfortunately, the license of this code is not clear -- for legal reasons we therefore assumed it to be closed source and applied black-box testing.

Dumping and examining the ETS system RAM revealed an array of function pointers at \texttt{0x3fffc000}, which turned out to be the kernel exception handler tables. Each entry of that table corresponds to one exception of the kernel exception class. The index of the table refers to the cause of the exception as described in the ISA manual.

Using this information, we overwrite the first entry of the exception table (the \texttt{IllegalInstructionCause}) with our custom exception handler function, thus ensuring execution of the \mechanism exception path (see \texttt{illegal\_instruction} in \autoref{cf_hijacking_fig}).

To be able to make use of the return path instrumentation, we need to ensure that our custom handler is registered before the first return of a protected function happens. Depending on which functions in the built firmware are instrumented, we searched for a way to move the hooking code to a different function for different firmware builds. This problem is similar to what led to the introduction of so-called Master Codes \cite{kenobi-gba} in cheat devices like Datel's Action Replay v3 for the GameBoy Advance. In essence, their system allowed end users to overwrite values in the games memory, thereby altering values like health or experience gained. To overwrite these values, a routine in the cheat modules ROM was used. The game-specific master code was then used to dynamically patch the original game's ROM to inject a jump to the cheat routine every few frames. Similar in spirit, our implementation allows the specification of a master function which will be hooked with the exception handler installation routine. For simplicity, we use the same hooking idea like in the call path instrumentation.

\subsection{Return Path Instrumentation}

\begin{figure}
	\begin{xtensar}
rsr.epc1 a3
L32R a2, canary
SUB a2, a2, a3
BEQZ a2, good
L32R a0, returnstack
ADDI.N a0, a0, -12
L32I.N a2, a0, 0
L32R a0, stack_chk_f
callx0 a0
good:
L32R a0, returnstack
ADDI.N a0, a0, -12
L32R a15, stack_ptr
ADDI.N a15, a15, 4
S32I.N a0, a15, 0
L32I.N a15, a0, 12
L32I.N a0, a0, 8
L32I.N a2, a1, 20
L32I.N a3, a1, 24
L32I.N a4, a1, 28
ADDI.N a1, a1, 256
wsr.epc1 a0
rfe
	\end{xtensar}
	\caption{Implementation of our illegal instruction exception handler in xtensa assembler.}
	\label{listing:hr-xtensa-handler}
\end{figure}

As described in \autoref{lbi}, each time a \mechanism-instrumented function returns, the processor generates an illegal instruction exception (see \autoref{cf_hijacking_fig}).
In this part, we describe how we handle these illegal instruction exceptions and outline a sample implementation of a stack corruption detection with \autoref{listing:hr-xtensa-handler}.

On the xtensa architecture, the address which triggered the exception is stored in a special register. We load this address and compare it to the \mechanism canary which we store in a fixed location in RAM. We only use the caller-saved registers \texttt{a2} to \texttt{a4} as these are saved on the stack by the calling function and can thus be utilized without affecting further execution.
This is reflected in lines 2-4 of \autoref{listing:hr-xtensa-handler}.

If the canary check fails, we do not return from the exception and may hence freely utilize any register. Because an overflow must have happened, those values are considered to be invalid in any case. As with traditional stack canaries, we invoke a special function \texttt{stack\_chk\_fail}, which handles the abort and dumping of the execution state (lines 5 to 8 in \autoref{listing:hr-xtensa-handler}). As this function is part of our instrumentation implementation, we chose a human-readable format that dumps all registers (except \texttt{a0}, which cannot be recovered), and roughly $384$ bytes of stack around the current stack address (\texttt{a1}).

In the good and usually executed path (starting from line 10), we restore all registers to their saved values. Subsequently, we ensure that the \mechanism is transparent to the instrumented software. We load registers \texttt{a2} to \texttt{a4} from the stack. We store register \texttt{a0}, the new return address, on the topmost cell(s) of our return stack. We adjust the stack pointer in \texttt{a1} to $\mathtt{a1}+0x100$ in order to restore the stack frame of the function.
Finally, we overwrite the special register containing the fault return address with the valid return address we saved when calling the instrumented function.

We now explore two possible applications of the linker instrumentation: execution tracing and detection of memory corruption.

\section{Evaluation} \label{sec:results}

In this section we evaluate the \mechanism using the example implementation presented in \autoref{sec:impl}. Specifically, we demonstrate that execution flow tracing and crash dump information extraction can be achieved using our method. Additionally, we investigate the implications of the \mechanism regarding size increase of the resulting binary and execution time. Finally, we showcase a fuzzing setup that relies on the \mechanism to collect crash information of a fuzzed \esp device.

For our tests, we implemented a simple xor-as-a-service test program, where anything that retrieved is byte-wise XORed with \texttt{0x42} and then sent back. It contains a stack-based buffer overflow vulnerability (see \autoref{listing:vuln}) to simulate a real-world scenario in which an attacker gains control of the device
by overwriting the saved return address on the stack. We then compiled this program using our call- and return path instrumentation.

\begin{figure}
	\begin{lstlisting}[language=c, breaklines=true, frame=single, numbers=left, xleftmargin=2em]
void ICACHE_FLASH_ATTR
shell_tcp_recvcb(void *arg, char *pusrdata, unsigned short length)
{
  struct espconn *pespconn = 
    (struct espconn *) arg;
  char   xorbuf[20];
  char   *x;

  ets_memcpy(xorbuf, pusrdata, length);
  ...
}
	\end{lstlisting}
	\caption{Excerpt of a vulnerable test program.}
	\label{listing:vuln}
\end{figure}

\subsection{Execution Tracing}
We utilized the call path instrumentation of the \mechanism to insert a dump function into every indirect function call of the object file that was compiled from our test
code. This dump function has the full program state at the time of the call available. Because any call to (UART) printf-functions that can print state information would clobber
any used registers, we took care to fully save (and restore) the contents of all registers. However, this is a necessity specific to the xtensa architecture, as its calling convention does not make use of any callee-save registers. Still, the needed stack space for these operations may limit the applicability of our dump function. Under tight memory constraints, e.g. when handling non-trivial recursion, our injected code could lower the limit of the maximum possible recursion count. For an example of the output of our test program, see lines 1 and 3 of \autoref{fig:dump-stack-entry}. We use this information to track values across functions and non-public API endpoints. This enables us to have a better understanding of the inner workings of the SDK besides the public reference documents.

\begin{figure}[t]
\begin{lstlisting}[breaklines=true, frame=single, numbers=right, linewidth=7.85cm]
(0x3ffe8070) a0=0x40229fb5 a15=0x3ffef500 name='tcpserver_connectcb' sp=3ffffd74 
tcp connection established
(0x3ffe8070) a0=0x4022a974 a15=0x0 name='shell_tcp_recvcb' sp=3ffffd84 
\end{lstlisting}
	\caption{Sample output of the call instrumentation of our test program. The first address is the current location of the topmost entry of the return stack used in return-path instrumentation.}
	\label{fig:dump-stack-entry}
\end{figure}

\begin{figure}[t]
	\begin{lstlisting}[breaklines=true, frame=single, numbers=right, linewidth=7.85cm]
*** STACK SMASH DETECTED***
returning from function shell_tcp_recvcb
halting execution. pc=23232328, canary=deaddead

Register state:
a0=(unk)        a4=00000000      
	a8=00000000    a12=00000000
a1=3ffffd90     a5=00000000      
	a9=00000000    a13=23232323
a2=000002d0     a6=00000008     
	a10=00000000    a14=23232323
a3=00000000     a7=46464646     
	a11=00000000    a15=40217868

stack dump at 3ffffd00:
0x3ffffd00: 00 00 00 00 23 23 23 23  00 00 00 00 98 fc ff 3f
...
0x3ffffd50: 23 23 23 23 23 23 23 23  a9 02 00 00 00 79 21 40
0x3ffffd60: 23 23 23 23 23 23 23 23  23 23 23 23 23 23 23 23
0x3ffffd70: 23 23 23 23 23 23 23 23  23 23 23 23 23 23 23 23
0x3ffffd80: 23 23 23 23 23 23 23 23  23 23 23 23 23 23 23 23
0x3ffffd90: 23 23 23 23 23 23 23 23  23 23 23 23 23 23 23 23
...
	\end{lstlisting}
	\caption{Example of crash dump information that can be extracted using the \mechanism.}
	\label{fig:exec-state}
\end{figure}

\subsection{Crash Dump Information}
We also utilized the return-path instrumentation to collect crash dump information if the saved return address does not match our canary value.
As pictured in \autoref{fig:exec-state}, we are able to recover and print all registers (except \texttt{a0}) and the stack frame of the faulty function. We triggered this dump by exploiting the test vulnerability through sending a large amount of \texttt{a} characters. We see that the stack pointer at the time of failure points to \texttt{0x3ffffd90}, and from the generated assembly we can deduce that our initial, overflown buffer, was located at \texttt{0x3ffffd60}. The return address was saved to \texttt{0x3ffffd8c} and was overwritten. Because we saved the function name that was called to a stack structure during a function call using our call-path instrumentation, we can identify the function that contained the fault.

\subsection{Binary Size and Performance Overhead}
\begin{figure}
	\begin{tikzpicture}
		\pgfplotstableread{ % Read the data into a table macro
		Name                Increase
		liblwip\_536.a       102.53
		libmain.a		    127.64
		libmbedtls.a         82.52
		libnet80211.a        71.65
		libphy.a            146.73
		libpp.a	            113.39
		libpwm.a	        116.98
		libsmartconfig.a     33.73
		libssl.a             68.73
		libupgrade.a         29.95
		libwpa.a             61.81
		libwpa2.a            57.47
		libwps.a             59.07
		libairkiss.a         77.00
		libat.a              72.10
		libc.a               11.53
		libcrypto.a          65.45
		libdriver.a         141.36
		libespnow.a         111.19
		libgcc.a              4.83
		libhal.a            140.30
		libjson.a           129.70
		liblwip.a           108.72
		}\datatable
	
		\begin{axis}[
			xbar,
			xmin=0,
			ytick=data,
			yticklabels from table={\datatable}{Name},
			y=0.3cm,
			bar width = 0.2cm,
			width=.8\columnwidth
		]
			\addplot [fill=gray!20] table [x=Increase, y expr=\coordindex] {\datatable};
		
		\end{axis}

	\end{tikzpicture}
	\caption{Increases in size in percentage of libraries instrumented with the \mechanism.}
	\label{fig:size-increase-chart}
\end{figure}
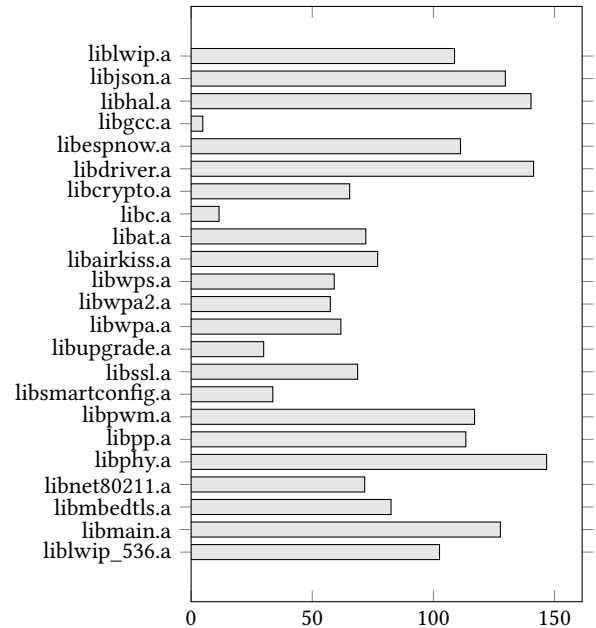

There is a non-trivial size increase when using the \mechanism; making complete
instrumentation of all SDK functions in real-world applications not practical.
However, the size increase varies greatly between the different libraries (see
\autoref{fig:size-increase-chart}), so it is possible to choose exactly those
of interest. Generally, a higher size increase is directly related to better
code coverage of the instrumentation. The best coverage can be achieved when
compiling with gcc's \texttt{-ffunction-sections} argument as each function is
placed into its own section, thus retaining a symbol name that can be
instrumented. In this scenario library-internal functions that are not part of
any exposed API can be fuzzed as well.

A special case considering size increases is \texttt{libgcc.a}. Although this library is relatively big (about $73$ KiB), it's size increase is only $4,87\%$, making it the library with the smallest size increase. This is due to the library mostly containing softmath library code which will not be instrumented at all.

Even with optimizations and hand-written assembly using narrow instruction encodings to reduce the size of the wrapping code, the introduced overhead is still large for embedded systems and especially the \esp. Currently, the SDK of our example platform supports only up to 16 Mbit SPI flash chips which is not large enough to instrument all SDK functions in a given application firmware at once.

As a result, we can only instrument parts of the firmware at one time when fuzzing, for instance. We automate the instrumentation process, automating the unpacking of the contents of a given archive (if necessary), renaming the symbols in the object files, and generating wrappers as described, compiling them and re-packing everything to an archive that can be used for linking. All this can be configured either in the projects Makefile or via environment variables for great flexibility.

The execution time overhead is not tied to the size increase in any form, as the size increase stems mainly from the addition of function-dependent hooks that get added to the archive. Instead, the overhead scales with the amount of code that is used in the instrumentation. In our case, the biggest time sink was the printing to UART using a rather low baud rate, which was fine for our test application. However, when instrumenting time-critical code in future work we need to be careful as to not break timings on e.g. the physical 802.11 layer.

\subsection{Fuzzing Setup} \label{sec:fuzz.setup}
\begin{figure*}[t]
 \def\svgwidth{.9\textwidth}
 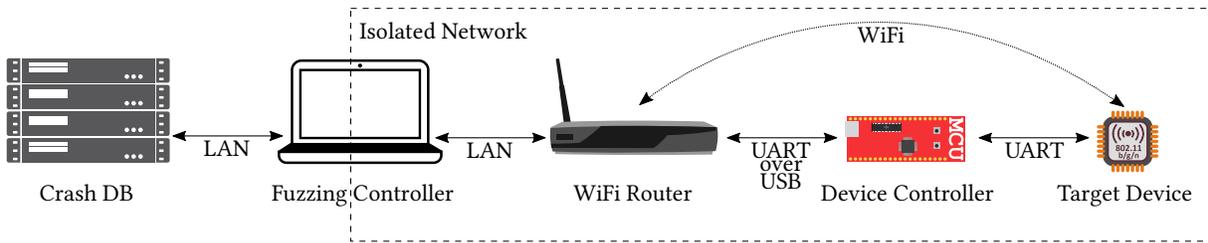
 \caption{A schematic drawing of our fuzzing setup.}
 \label{fig:fuzzing_setup}
\end{figure*}

In general, there are three modes of operation supported by the \esp: Access Point, Station and Mesh. 
For our tests, we wanted to mimic a typical household \gls{IoT}-Scenario. We deployed a common off 
the shelf router running OpenWRT to which the target \esp in station mode connects. We connected the host device, a desktop computer, running the fuzzer via a standard Ethernet connection to the router. The whole setup is depicted in \autoref{fig:fuzzing_setup}.

Our fuzzer is based on BooFuzz \cite{boofuzz-github}, a popular python layer 2/3 protocol-based modelling fuzzer. 
BooFuzz is capable of fuzzing various protocols of the \esp's network stack.
For our testing purposes and to show the effectiveness of the \mechanism, we built a sample vulnerable binary outlined
in \autoref{listing:vuln}. This binary contains a vulnerability that must be found while fuzzing with the \mechanism.

We used our return-path instrumenting to print crash output and memory dumps from the \esp to the UART serial connection. For efficient fault capturing, we need to capture the dumped information. Additionally, we must ensure that we can reboot the \esp from every state, even if it is completely hung up, without human interaction (\eg pulling the power cable).

To address both problems, we used a target device controller board, in our setup a SparkFun ESP32thing, that captures the UART output of the \esp and forwards it to its own serial connection with the host computer. Additionally, this board may reset the \esp on command by pulling its \texttt{RST} pin to \texttt{GND}. The controller board also provides power to the \esp.

To aid debugging of the whole setup, we multiplex the serial connection of the ESP32 to $n$ network connections to provide
an easy to use back-channel to the fuzzer while being able to monitor the serial output independently.

\subsection{Practical Attacks on the \esp}
For the following, we categorize our efforts in two categories: First, we sketch a method to achieve chosen payload execution, second, we discuss how we approach a permanent denial of service attack.

Identifying exploitable vulnerabilities in \esp firmware might quickly lead to severe consequences.
Code execution can be gained either by a traditional buffer-overflow based attack, whether on the stack, the heap or in a structure in a R/W mapped data section. Additionally, the \esp provides an easy-to-use firmware update mechanism. However, there is absolutely no protection against corrupting the \esp's firmware; even the download has to be 
done via unencrypted HTTP as HTTPS is not supported. Each firmware contains a 16 bit integer describing a version number. The new, downloaded firmware that is to be flashed onto the flash memory is then checked if its version number is greater than the one that is currently booted. If so,
the download continues and the flash gets overwritten with whatever is presented to the update mechanism. Obviously, setting
the version number to $2^{16}-1$ disables the update mechanism until the device is retrieved and manually flashed via a UART
download. We can also achieve the latter by directly overwriting the version field present in the SPI flash, bypassing the update mechanism.

Another method to permanently put an \esp out of order is by physically damaging the flash chip through excessively reading and writing single flash cells in the image header region, thus hindering boot. Templeman and Kapadia introduced GANGRENE \cite{gangrene} that shows the feasibility of such an attack.

\section{Discussion}
In the following we discuss important aspects and limitations of our method in general as well as specifics of our implementation.

The recoverability of an attack of the \esps firmware greatly depends on the specific deployment of the device. In home \gls{IoT} settings it is
hardly imaginable that such a device would be user-serviceable given the current lax standards regarding (security) updates in
consumer devices in the first place. In a more professional environment it is possible to re-flash a fixed version of the firmware
manually to a limited amount of devices provided the OEM designed the specific application in a way a UART download flash is possible.
To go to extreme lengths, devices could also be recovered by just swapping out the SPI flash chip with a known good one; albeit that
this would involve some soldering. However, in a mesh network with a large number of deployed devices, recovering them most likely is not an option
if, for example, the exact location of the devices is unknown due to aerial deployment. 

While we managed to achieve our goal providing more and clearer output at the point of failure, our method obviously has a few limitations.
First and foremost, while we provide a valuable tool for fuzzing, the \mechanism is not designed to be a security feature. It relies on the fact that
the return of a function generates an exception that we can catch. An attacker would simply be able to overwrite the saved return pointer on
the stack and gain control of the execution flow, completely bypassing the \mechanism.

It is unlikely although possible that automated software testing overwrites the saved return address in a way no exception gets triggered, which would
corrupt the return stack on subsequent calls. This also holds true for accidentally writing the canary value to the saved return address when in fact
an overwrite did happen. Such an access would not be detectable by our implementation of the return address checker.

Obviously there is also no generic function for dumping memory contents in the case of failure. Even with limited amounts of RAM the
address space is only sparsely mapped, so the dumping function needs to be aware of the target chips memory map. In some cases,
it may be even specific to the SDK used for development of the chip firmware, as information about \eg the heap usage could
be directly displayed in the dump. Still, this enables researchers to more easily obtain information about a chips internal state
even when no debug interfaces like JTAG are available.

Recently, \gls{IoT} devices with multiple cores appeared on the market. While our example implementation is tied to the single-core \esp board, our method is applicable to systems employing parallel execution as well. For the call-path instrumentation nothing needs to be changed. The return-path instrumentation hook however needs to maintain a distinct return stack for each execution strain (usually each core).

As it is the case with the \esp, oftentimes storage size restrictions prevent instrumenting of the whole instrumented firmware due
to the added overhead. While SPI flash memory is cheaply available online, not every board supports memory-mapping SPI flash chips of
arbitrary size. A workaround to this may be to employ some kind of bank-switching scheme where the firmware switches between SPI chips at runtime.
The feasibility of this, especially considering interrupt handling routines, remains to be researched in future work.

In some scenarios, \eg smart door locks or industrial applications, it might be needed to connect
actuators to the chip that are not easily moved. Our setup described in \autoref{sec:fuzz.setup} allows to easily separate the fuzzer host 
and the fuzzed board, making fuzzing easier while the device is deployed. 

\section{Related Work}
In this section we cover related work in embedded (static) binary instrumentation and software testing.

Muench, et al. \cite{EURECOM+5417} pointed out that memory corruptions in embedded devices oftentimes result in
different behavior than in desktop systems. The \esp, according to their paper, is a Type III device, with a single monolithic firmware model and no OS.
Muench et al. observed no visible crashes (or reboots) while probing their Type III device. The \mechanism aims to improve this situation through e.g. our stack overflow detection as described in \autoref{sec:impl}.

Thomas' LIEF project \cite{LIEF} is like the \mechanism a framework to instrument binary files. It handles substantially more formats than our work, but is mainly designed for
 systems that have an OS (Type I or Type II according to \cite{EURECOM+5417}). While it can rewrite parts of an ELF file, we found this capability rather unstable on uncommon architectures like xtensa. Because LIEF only handles
files in a commonly known executable format, it cannot process flat firmware images for development boards out of the box. As is the case with the \mechanism, LIEF therefore is only applicable before linking as libraries in the \esps SDK are provided in ELF format. In particular, we used the symbol parsing part of LIEF for the \mechanism.

Corteggiani, Camurati and Francillon \cite{217620} introduce Inception, a framework for symbolic execution of embedded systems software. While able to operate on binaries without available source code, it still requires a JTAG port present on the target device. Avoiding this was an explicit design goal of the \mechanism.

Song et al. \cite{song-ndss-2019} published PeriScope, a probing and fuzzing framework for the hardware-OS boundary. While interested in a related target, network stacks, their approach relies on the presence of an MMU to intercept memory access. Because on many embedded systems no memory management is available, we designed the \mechanism in a way that works without one. However, this limits the effectiveness of our method compared to PeriScope.

\section{Conclusion}
Motivated by the rapidly growing distribution of heavily interconnected embedded devices, we proposed the \mechanism, a method for embedded firmware testing.
The \mechanism is especially useful for security testing of \gls{IoT} deployments using closed-source firmware components, which can potentially introduce fatal vulnerabilities.
With the \mechanism, we hook the control flow of firmware on calls and returns to and from subroutines.
This allows us fine-grained insight to code execution flow and to detect stack overflows.
While keeping the design of the \mechanism independent of embedded architectures, we implemented a prototype for the xtensa architecture.
Our instrumentation method for libraries from any archive is ELF specific, but generally architecturally independent.
We evaluated the usefulness at the example of the \esp Wi-Fi chip, showcasing the tracing of execution flows and the detection of stack-based buffer overflows.
The \mechanism can easily be ported to different chips and architectures, as long as exception handling is available.
Depending on the architecture specifics, the overhead may be significantly less or more than what was observed on the \esp platform.
Furthermore, our setup for the \esp can easily be adapted for various other embedded scenarios and be used for wireless protocol fuzzing such as bluetooth or Wi-Fi protocol implementations.
This allows the identification of flaws in protocol implementations, reducing the risk of the exploitation of a large number of devices sharing the same vulnerable library implementations.
We aim to open source our implementation shortly after the publication of the paper at the Fraunhofer AISEC GitHub page: \href{https://github.com/Fraunhofer-AISEC}{https://github.com/Fraunhofer-AISEC}.

%acm
\bibliographystyle{ACM-Reference-Format}
\bibliography{biblio}


\begin{thebibliography}{24}

%%% the Web address.

\ifx \showCODEN    \undefined \def \showCODEN     #1{\unskip}     \fi
\ifx \showDOI      \undefined \def \showDOI       #1{#1}\fi
\ifx \showISBNx    \undefined \def \showISBNx     #1{\unskip}     \fi
\ifx \showISBNxiii \undefined \def \showISBNxiii  #1{\unskip}     \fi
\ifx \showISSN     \undefined \def \showISSN      #1{\unskip}     \fi
\ifx \showLCCN     \undefined \def \showLCCN      #1{\unskip}     \fi
\ifx \shownote     \undefined \def \shownote      #1{#1}          \fi
\ifx \showarticletitle \undefined \def \showarticletitle #1{#1}   \fi
\ifx \showURL      \undefined \def \showURL       {\relax}        \fi
% invisible to TeX
\providecommand\bibfield[2]{#2}
\providecommand\bibinfo[2]{#2}
\providecommand\natexlab[1]{#1}
\providecommand\showeprint[2][]{arXiv:#2}

\bibitem[\protect\citeauthoryear{Corporation}{Corporation}{2015}]%
        {winbond-flash}
\bibfield{author}{\bibinfo{person}{Winbond~Electronics Corporation}.}
  \bibinfo{year}{2015}\natexlab{}.
\newblock \bibinfo{booktitle}{\emph{Winbond W25Q128FV Datasheet}}.
\newblock
\urldef\tempurl%
\url{https://www.winbond.com/resource-files/w25q128fv%20rev.l%2008242015.pdf}
\showURL{%
\tempurl}


\bibitem[\protect\citeauthoryear{Corteggiani, Camurati, and
  Francillon}{Corteggiani et~al\mbox{.}}{2018}]%
        {217620}
\bibfield{author}{\bibinfo{person}{Nassim Corteggiani},
  \bibinfo{person}{Giovanni Camurati}, {and} \bibinfo{person}{Aur{\'e}lien
  Francillon}.} \bibinfo{year}{2018}\natexlab{}.
\newblock \showarticletitle{Inception: System-Wide Security Testing of
  Real-World Embedded Systems Software}. In \bibinfo{booktitle}{\emph{27th
  {USENIX} Security Symposium ({USENIX} Security 18)}}.
  \bibinfo{publisher}{{USENIX} Association}, \bibinfo{address}{Baltimore, MD},
  \bibinfo{pages}{309--326}.
\newblock
\showISBNx{978-1-939133-04-5}
\urldef\tempurl%
\url{https://www.usenix.org/conference/usenixsecurity18/presentation/corteggiani}
\showURL{%
\tempurl}


\bibitem[\protect\citeauthoryear{dgtlrift}{dgtlrift}{2016}]%
        {esp-qemu}
\bibfield{author}{\bibinfo{person}{dgtlrift}.} \bibinfo{year}{2016}\natexlab{}.
\newblock \bibinfo{booktitle}{\emph{Patch: QEMU simulation of ESP8266 prior to
  flashing}}.
\newblock
\urldef\tempurl%
\url{https://github.com/SuperHouse/esp-open-rtos/issues/230}
\showURL{%
\tempurl}


\bibitem[\protect\citeauthoryear{Eduardo}{Eduardo}{2019}]%
        {esp-hacks}
\bibfield{author}{\bibinfo{person}{Matheus Eduardo}.}
  \bibinfo{year}{2019}\natexlab{}.
\newblock \bibinfo{booktitle}{\emph{Proof of Concept of ESP32/8266 Wi-Fi
  vulnerabilties (CVE-2019-12586, CVE-2019-12587, CVE-2019-12588)}}.
\newblock
\urldef\tempurl%
\url{https://github.com/Matheus-Garbelini/esp32_esp8266_attacks}
\showURL{%
\tempurl}


\bibitem[\protect\citeauthoryear{et.al.}{et.al.}{[n.d.]}]%
        {boofuzz-github}
\bibfield{author}{\bibinfo{person}{Joshua~Pereyda et.al.}}
  \bibinfo{year}{[n.d.]}\natexlab{}.
\newblock \bibinfo{booktitle}{\emph{BooFuzz Source Code repository}}.
\newblock
\urldef\tempurl%
\url{https://github.com/jtpereyda/boofuzz}
\showURL{%
\tempurl}


\bibitem[\protect\citeauthoryear{et.al}{et.al}{2015}]%
        {esp-memory-map}
\bibfield{author}{\bibinfo{person}{Max~Filippov et.al}.}
  \bibinfo{year}{2015}\natexlab{}.
\newblock \bibinfo{booktitle}{\emph{esp8266 Memory Map}}.
\newblock
\urldef\tempurl%
\url{https://github.com/esp8266/esp8266-wiki/wiki/Memory-Map}
\showURL{%
\tempurl}


\bibitem[\protect\citeauthoryear{Filippov}{Filippov}{2015}]%
        {crosstool-esp-config}
\bibfield{author}{\bibinfo{person}{Max Filippov}.}
  \bibinfo{year}{2015}\natexlab{}.
\newblock \bibinfo{booktitle}{\emph{esp8266 processor feature config}}.
\newblock
\urldef\tempurl%
\url{https://github.com/jcmvbkbc/crosstool-NG/blob/xtensa-1.22.x/overlays/xtensa_lx106.tar#L16085}
\showURL{%
\tempurl}


\bibitem[\protect\citeauthoryear{Inc.}{Inc.}{2019}]%
        {xtensa-isa}
\bibfield{author}{\bibinfo{person}{Tensilica Inc.}}
  \bibinfo{year}{2019}\natexlab{}.
\newblock \bibinfo{booktitle}{\emph{Xtensa\textsuperscript{\textregistered}
  Instruction Set Architecture (ISA) Reference Manual}}.
\newblock
\urldef\tempurl%
\url{https://0x04.net/~mwk/doc/xtensa.pdf}
\showURL{%
\tempurl}


\bibitem[\protect\citeauthoryear{Kammerstetter, Platzer, and
  Kastner}{Kammerstetter et~al\mbox{.}}{2014}]%
        {Kammerstetter:2014:PPP:2590296.2590301}
\bibfield{author}{\bibinfo{person}{Markus Kammerstetter},
  \bibinfo{person}{Christian Platzer}, {and} \bibinfo{person}{Wolfgang
  Kastner}.} \bibinfo{year}{2014}\natexlab{}.
\newblock \showarticletitle{Prospect: Peripheral Proxying Supported Embedded
  Code Testing}. In \bibinfo{booktitle}{\emph{Proceedings of the 9th ACM
  Symposium on Information, Computer and Communications Security}}
  \emph{(\bibinfo{series}{ASIA CCS '14})}. \bibinfo{publisher}{ACM},
  \bibinfo{address}{New York, NY, USA}, \bibinfo{pages}{329--340}.
\newblock
\showISBNx{978-1-4503-2800-5}
\urldef\tempurl%
\url{https://doi.org/10.1145/2590296.2590301}
\showDOI{\tempurl}


\bibitem[\protect\citeauthoryear{{Kolias}, {Kambourakis}, {Stavrou}, and
  {Voas}}{{Kolias} et~al\mbox{.}}{2017}]%
        {mirai-iot-attacks}
\bibfield{author}{\bibinfo{person}{Constantinos {Kolias}},
  \bibinfo{person}{Georgios {Kambourakis}}, \bibinfo{person}{Angelos
  {Stavrou}}, {and} \bibinfo{person}{Jeffrey {Voas}}.}
  \bibinfo{year}{2017}\natexlab{}.
\newblock \showarticletitle{DDoS in the IoT: Mirai and Other Botnets}.
\newblock \bibinfo{journal}{\emph{Computer}} \bibinfo{volume}{50},
  \bibinfo{number}{7} (\bibinfo{year}{2017}), \bibinfo{pages}{80--84}.
\newblock
\urldef\tempurl%
\url{https://doi.org/10.1109/MC.2017.201}
\showDOI{\tempurl}


\bibitem[\protect\citeauthoryear{LTD.}{LTD.}{2017}]%
        {sdk-license}
\bibfield{author}{\bibinfo{person}{Espressif Systems~CO. LTD.}}
  \bibinfo{year}{2017}\natexlab{}.
\newblock \bibinfo{booktitle}{\emph{Espressif MIT License}}.
\newblock
\urldef\tempurl%
\url{https://github.com/espressif/ESP8266_NONOS_SDK/blob/90c641efe84066b47c4616ed367697a9f49f3ac5/License}
\showURL{%
\tempurl}


\bibitem[\protect\citeauthoryear{LTD.}{LTD.}{2019}]%
        {sdk-github}
\bibfield{author}{\bibinfo{person}{Espressif Systems~CO. LTD.}}
  \bibinfo{year}{2019}\natexlab{}.
\newblock \bibinfo{booktitle}{\emph{ESP8266 NONOS SDK Source Code}}.
\newblock
\urldef\tempurl%
\url{https://github.com/espressif/ESP8266_NONOS_SDK}
\showURL{%
\tempurl}


\bibitem[\protect\citeauthoryear{{M}uench, {S}tijohann, {K}argl, {F}rancillon,
  and {B}alzarotti}{{M}uench et~al\mbox{.}}{2018}]%
        {EURECOM+5417}
\bibfield{author}{\bibinfo{person}{{M}arius {M}uench}, \bibinfo{person}{{J}an
  {S}tijohann}, \bibinfo{person}{{F}rank {K}argl},
  \bibinfo{person}{{A}ur{\'e}lien {F}rancillon}, {and}
  \bibinfo{person}{{D}avide {B}alzarotti}.} \bibinfo{year}{2018}\natexlab{}.
\newblock \showarticletitle{{W}hat you corrupt is not what you crash:
  {C}hallenges in fuzzing embedded devices}. In
  \bibinfo{booktitle}{\emph{{NDSS} 2018, {N}etwork and {D}istributed {S}ystems
  {S}ecurity {S}ymposium, 18-21 {F}ebruary 2018, {S}an {D}iego, {CA}, {USA}}}.
  \bibinfo{address}{{S}an {D}iego, {UNITED} {STATES}}.
\newblock
\urldef\tempurl%
\url{http://www.eurecom.fr/publication/5417}
\showURL{%
\tempurl}


\bibitem[\protect\citeauthoryear{Offermanns}{Offermanns}{2015}]%
        {doi:10.1002/ciuz.201500708}
\bibfield{author}{\bibinfo{person}{Heribert Offermanns}.}
  \bibinfo{year}{2015}\natexlab{}.
\newblock \showarticletitle{Kanarienvögel und Grubenlampen}.
\newblock \bibinfo{journal}{\emph{Chemie in unserer Zeit}}
  \bibinfo{volume}{49}, \bibinfo{number}{2} (\bibinfo{year}{2015}),
  \bibinfo{pages}{140--141}.
\newblock
\urldef\tempurl%
\url{https://doi.org/10.1002/ciuz.201500708}
\showDOI{\tempurl}
\showeprint{https://onlinelibrary.wiley.com/doi/pdf/10.1002/ciuz.201500708}


\bibitem[\protect\citeauthoryear{Rool, Macrox, Tolos, DGenerateKane,
  HyperHacker, Viper187, and Kenobi}{Rool et~al\mbox{.}}{2004}]%
        {kenobi-gba}
\bibfield{author}{\bibinfo{person}{Kong~K Rool}, \bibinfo{person}{Macrox},
  \bibinfo{person}{Tolos}, \bibinfo{person}{DGenerateKane},
  \bibinfo{person}{HyperHacker}, \bibinfo{person}{Viper187}, {and}
  \bibinfo{person}{Kenobi}.} \bibinfo{year}{2004}\natexlab{}.
\newblock \bibinfo{booktitle}{\emph{The Secrets of Professional Gameshark(tm)
  Hacking}}.
\newblock
\urldef\tempurl%
\url{https://macrox.gshi.org/The%20Hacking%20Text.htm#gba_non_standard_master}
\showURL{%
\tempurl}


\bibitem[\protect\citeauthoryear{Sokolovsky}{Sokolovsky}{2019}]%
        {esp-open-sdk}
\bibfield{author}{\bibinfo{person}{Paul Sokolovsky}.}
  \bibinfo{year}{2019}\natexlab{}.
\newblock \bibinfo{booktitle}{\emph{ESP Open SDK}}.
\newblock
\urldef\tempurl%
\url{https://github.com/pfalcon/esp-open-sdk}
\showURL{%
\tempurl}


\bibitem[\protect\citeauthoryear{Song, Hetzelt, Das, Spensky, Na, Volckaert,
  Vigna, Kruegel, Seifert, and Franz}{Song et~al\mbox{.}}{2019}]%
        {song-ndss-2019}
\bibfield{author}{\bibinfo{person}{Dokyung Song}, \bibinfo{person}{Felicitas
  Hetzelt}, \bibinfo{person}{Dipanjan Das}, \bibinfo{person}{Chad Spensky},
  \bibinfo{person}{Yeoul Na}, \bibinfo{person}{Stijn Volckaert},
  \bibinfo{person}{Giovanni Vigna}, \bibinfo{person}{Christopher Kruegel},
  \bibinfo{person}{Jean-Pierre Seifert}, {and} \bibinfo{person}{Michael
  Franz}.} \bibinfo{year}{2019}\natexlab{}.
\newblock \showarticletitle{PeriScope: An Effective Probing and Fuzzing
  Framework for the Hardware-OS Boundary}. In
  \bibinfo{booktitle}{\emph{{Proceedings of the Network and Distributed System
  Security Symposium (NDSS 2019)}}}. \bibinfo{address}{San Diego, CA}.
\newblock


\bibitem[\protect\citeauthoryear{systems CO.~LTD.}{systems CO.~LTD.}{2017a}]%
        {esp-exception-ld}
\bibfield{author}{\bibinfo{person}{Espressif systems CO.~LTD.}}
  \bibinfo{year}{2017}\natexlab{a}.
\newblock \bibinfo{booktitle}{\emph{esp8266 ROM addresses: Exception Vectors}}.
\newblock
\urldef\tempurl%
\url{https://github.com/espressif/ESP8266_NONOS_SDK/blob/release/v3.0.0/ld/eagle.rom.addr.v6.ld#L45}
\showURL{%
\tempurl}


\bibitem[\protect\citeauthoryear{systems CO.~LTD.}{systems CO.~LTD.}{2017b}]%
        {esp-sales}
\bibfield{author}{\bibinfo{person}{Espressif systems CO.~LTD.}}
  \bibinfo{year}{2017}\natexlab{b}.
\newblock \bibinfo{booktitle}{\emph{Espressif Achieves 100 Million Target in
  IoT Chip Shipments}}.
\newblock
\urldef\tempurl%
\url{https://www.espressif.com/en/media_overview/news/espressif-achieves-100-million-target-iot-chip-shipments}
\showURL{%
\tempurl}


\bibitem[\protect\citeauthoryear{systems CO.~LTD.}{systems CO.~LTD.}{2019a}]%
        {esp32-datasheet}
\bibfield{author}{\bibinfo{person}{Espressif systems CO.~LTD.}}
  \bibinfo{year}{2019}\natexlab{a}.
\newblock \bibinfo{booktitle}{\emph{ESP32 Datasheet}}.
\newblock
\urldef\tempurl%
\url{https://www.espressif.com/sites/default/files/documentation/esp32_datasheet_en.pdf}
\showURL{%
\tempurl}


\bibitem[\protect\citeauthoryear{systems CO.~LTD.}{systems CO.~LTD.}{2019b}]%
        {esp-datasheet}
\bibfield{author}{\bibinfo{person}{Espressif systems CO.~LTD.}}
  \bibinfo{year}{2019}\natexlab{b}.
\newblock \bibinfo{booktitle}{\emph{ESP8266 Datasheet}}.
\newblock
\urldef\tempurl%
\url{https://www.espressif.com/sites/default/files/documentation/0a-esp8266ex_datasheet_en.pdf}
\showURL{%
\tempurl}


\bibitem[\protect\citeauthoryear{systems CO.~LTD.}{systems CO.~LTD.}{2019c}]%
        {esp8266-sdk-reference}
\bibfield{author}{\bibinfo{person}{Espressif systems CO.~LTD.}}
  \bibinfo{year}{2019}\natexlab{c}.
\newblock \bibinfo{booktitle}{\emph{ESP8266 Non-OS SDK API Reference}}.
\newblock
\urldef\tempurl%
\url{https://www.espressif.com/sites/default/files/documentation/2c-esp8266_non_os_sdk_api_reference_en.pdf}
\showURL{%
\tempurl}


\bibitem[\protect\citeauthoryear{Templeman and Kapadia}{Templeman and
  Kapadia}{2012}]%
        {gangrene}
\bibfield{author}{\bibinfo{person}{Robert Templeman} {and} \bibinfo{person}{Apu
  Kapadia}.} \bibinfo{year}{2012}\natexlab{}.
\newblock \showarticletitle{{GANGRENE}: Exploring the Mortality of Flash
  Memory}. In \bibinfo{booktitle}{\emph{Presented as part of the 7th {USENIX}
  Workshop on Hot Topics in Security}}. \bibinfo{publisher}{{USENIX}},
  \bibinfo{address}{Bellevue, WA}.
\newblock
\urldef\tempurl%
\url{https://www.usenix.org/conference/hotsec12/workshop-program/presentation/Templeman}
\showURL{%
\tempurl}


\bibitem[\protect\citeauthoryear{Thomas}{Thomas}{2017}]%
        {LIEF}
\bibfield{author}{\bibinfo{person}{Romain Thomas}.}
  \bibinfo{year}{2017}\natexlab{}.
\newblock \bibinfo{title}{LIEF - Library to Instrument Executable Formats}.
\newblock \bibinfo{howpublished}{https://lief.quarkslab.com/}.
\newblock


\end{thebibliography}

\end{document}